\documentstyle[12pt]{article}
\newcommand{\be}{\begin{equation}}
\newcommand{\ee}{\end{equation}}
\newcommand{\bea}{\begin{eqnarray}}
\newcommand{\eea}{\end{eqnarray}}
\newcommand{\cl}{\centerline}

\def\anf#1{\hbox{,\kern -0.80pt,\kern 0.3pt}#1\kern -0.7pt``}

\def\no#1{\hbox{\kern 0.3em}#1\hspace{-0.6em}/}
\def\exp{\mbox{\hspace{1mm} exp}}
\def\expo#1{\exp\left\{#1 \right\}}
\def\tr{\mbox{tr}}
\def\pppp#1{\frac{\dddd \mbox{\hspace{1mm}} #1}{(2 \pi)^4}}

\def\pp#1{\frac{\dd \mbox{\hspace{1mm}} #1}{(2 \pi)^2}}

\newcommand{\mspace}{\mbox{\hspace{1mm}}}
\newcommand{\dddd}{{\rm d}^4\!}

\newcommand{\dd}{{\rm d}^2\!}
\newcommand{\de}{{\rm d}\mspace \!}

\newcommand{\GeV}{{\,\mbox{GeV}}}

\newcommand{\pmb}[1]{%
\setbox0=\hbox{#1}%
\kern-.02em\copy0\kern-\wd0
\kern+.04em\copy0\kern-\wd0
\kern-.02em\raise.0217em\box0}
\newcommand{\vek}[1]{
\mathchoice{\mbox{\boldmath$#1$}}%
{\mbox{\boldmath$#1$}}%
{\pmb{$\scriptstyle#1$}}%
{\pmb{$\scriptscriptstyle#1$}}}

\begin{document}

\thispagestyle{empty}

\renewcommand{\thefootnote}{\fnsymbol{footnote}}

\parindent0cm

\rule[-0.21cm]{\textwidth}{0.05cm}\\ \rule[0.15cm]{\textwidth}{0.01cm}

March, 1995\hfill TUM/T39-95-3

\rule[0.1cm]{\textwidth}{0.01cm}\\ \rule[0.75cm]{\textwidth}{0.05cm}

\vspace{1cm}

{\Large\bf{ \cl{NON-PERTURBATIVE SCALES  }
           \cl{IN SOFT HADRONIC COLLISIONS}
           \cl{AT HIGH ENERGY$^{\ast )}$}
}

\vspace{2cm}

\normalsize

\centerline{U. Grandel$^{\dagger )}$ and W. Weise}
\vspace{0.1cm}
\centerline{\small Physik-Department}
\centerline{\small Technische Universit\"at M\"unchen}
\centerline{\small Institut f\"ur Theoretische Physik}
\centerline{\small  D-85747 Garching,  Germany}
\vspace{0.5cm}

\vspace{1.5cm}

\begin{abstract}

We investigate the role of nonperturbative quark-gluon dynamics in soft
high energy processes. In order to reproduce differential and total
cross sections
for elastic proton-proton and proton-antiproton-scattering at high energy
and small momentum transfer it turns out that we need two
scales, the gluonic correlation length and a confinement scale.
We find a small gluonic correlation length, $a \simeq 0.2$ fm, in accordance
with recent lattice QCD results.

\end{abstract}
}
\vspace{2cm}

\begin{center}

{\sl Submitted to Phys. Lett. B}

\end{center}

\vfill

\rule{5cm}{0.5mm}

$^{\ast )}$ Work supported in part by BMFT.\\
$^{\dagger )}$ Supported in part by  Hans Boeckler and Friedrich Ebert
 Foundations.

\clearpage

{\bf Introduction.}
Soft hadronic collisions at high energy are
described successfully using Regge and Pomeron phenomenology \cite{1,2}.
Yet an outstanding problem is the more basic understanding of pomeron
exchange in terms of the underlying quark-gluon-dynamics.

\medskip

At large center of mass energies $\sqrt{s}$ but small momentum transfers
$|t| \stackrel{<}{\sim} (1 - 2) $ GeV$^2$, one expects that
non-perturbative features of QCD, such as gluonic vacuum properties,
 enter as important ingredients in hadron-hadron scattering amplitudes.
In this context several recent papers \cite{3,4,5} have emphasized
specifically the role of gluonic field strength correlations in hadronic
scattering processes at high energy.

\medskip

The aim of the present paper is to follow up on this theme and gain further
insight into the basic scales which govern soft high energy collisions.
In particular, the question arises whether the small gluonic correlation
length recently found in lattice QCD calculations \cite{6} is
compatible with the scales deduced from the $t$-dependence of differential
cross sections for high energy proton-proton and proton-antiproton
collisions.
Our starting point is the assumption that the scattering of composite
hadrons can be reduced to the scattering of their quark constituents;
we develop  a model of quark-quark scattering  in terms of
non-perturbative
gluon dynamics. Our basic picture is that soft interactions
between quarks at high energy are described by the propagation
of projectile quarks in a nonperturbative gluonic environment
produced by the target quarks. The quark sources themselves
are tied  together by confining forces inside the hadron projectile
and target.

\bigskip

{\bf Propagation of High Energy Quarks in a Gluonic Background.}
We start from the QCD Lagrangian

\begin{equation}
{\cal L}(x) =  -\frac{1}{2} \tr G_{\mu\nu}(x) G^{\mu\nu}(x) +
\bar{\psi}(x) \left[ i\gamma ^{\mu} D_{\mu}(x) - m_0 \right]
\psi(x)
\end{equation}

with
the usual field strength tensor $G_{\mu\nu} = \partial_{\mu} G_{\nu} -
\partial_{\nu} G_{\mu} + i g [G_{\mu}, G_{\nu}]$ and the covariant

derivative  $D_{\mu}(x) = \partial_{\mu} + i g G_{\mu}(x)$.
Here  $G_{\mu}(x) = G_{\mu}^a(x) T_a $ are the gluon fields
contracted with the $SU(3)_{\rm color}$ generators $T_a = \lambda_a/2$.
The quark field $\psi(x)$ includes three color components, and $m_0$ is the
current quark mass.
Consider the vacuum-vacuum transition amplitude

\begin{equation}
{\cal Z}
= \int {\cal D} G \mspace {\cal D} \psi
\mspace {\cal D} \bar{\psi} \mspace
\expo{i \int \dddd x \mspace  {\cal L}(x)} ,
\end{equation}

where the integration is over all independent quark and gluon fields.
Our aim is to reduce the complexity of the gluon field configurations
appearing in eq. $(2)$ to the point where one can see the explicit appearance
of gluonic correlation functions.

\medskip

 Let $f[G, \psi]$ be a functional
of the gluon and quark fields.
We define an average over the gluon fields ($G$-average) as

\begin{equation}
\left< f[G, \psi] \right>_G :=  \int {\cal D} G  \mspace
\exp\left\{i\int \dddd x  \left[ -\frac{1}{2} \tr G_{\mu\nu}(x)
G^{\mu\nu}(x)\right] \right\} f[G, \psi] .
\end{equation}

For a pure Yang-Mills theory this is simply the expectation value of
$f[G]$. With the help of eq. $(3)$ the amplitude $(2)$ can be rewritten:

\begin{eqnarray}
{\cal Z}
&=& \int {\cal D} \psi
\mspace {\cal D} \bar{\psi} \mspace
\expo{ i \int \dddd x  \bar{\psi}(x) \left[ i\gamma ^{\mu}
\partial_{\mu} - m_0 \right] \psi(x)}  \cdot
\nonumber\\
&& \cdot  \left< {\cal P} \exp\left\{- i g \int \dddd x
j^{\mu}_a(x) G_{\mu}^a(x)
\right\}\right>_G , \nonumber\\
\end{eqnarray}

where $j^{\mu}_a (x) = \bar{\psi} (x) \gamma^{\mu} T_a \psi (x)$
is the quark  color current. Here ${\cal P}$ denotes Dyson
pathordering\cite{6a}.

In the next step we use the cumulant expansion \cite{7} to reexpress
the $G$-average  in $(4)$  as follows:

\begin{eqnarray}
\left< {\cal P} \exp\left\{-i g \int \dddd x  j^{\mu}_a
G^a_{\mu}\right\} \right>_G
&\simeq & \nonumber\\
\exp \{\sum_{m=1}^{\infty} \frac{(-i g)^m}{m!} \int \dddd x_1 \dots
\int \dddd x_m  && \mbox{\hspace{-1cm}}  j^{\mu _1}_{a_1} (x_1)
 \dots  j^{\mu _m}_{a_m} (x_m)
< G^{a_1}_{\mu _1}(x_1)  \dots  G^{a_m}_{\mu _m}(x_m) >_c \}.
\nonumber\\
\end{eqnarray}

The index $c$ now includes
Dyson-ordered cumulants. For a given function $f$, these are constructed out
of the $G$-averaged moments by collecting the connected parts:

\begin{eqnarray}
<f(x)>_c & = & <f(x)>_G , \nonumber\\
< f(x_1) f(x_2)>_c & = &
<f(x_1) f(x_2)>_G -  <f(x_1)>_G <f(x_2)>_G , \mbox{etc. }
\nonumber\\
\end{eqnarray}

Eq. $(5)$ is exact for commuting fields and correct
for noncommuting gluon fields if one includes moments up to
second order. For higher orders eq. $(5)$ is still a good approximation
if the  length scale of the underlying correlations is small \cite{8}
compared to the size of the interacting composite particles.

\medskip

We abbreviate the quark-gluon interaction part on the right hand side
of the cumulant expansion $(5)$ by $-ig \int \dddd x
j^{\mu}_{a} (x) < \tilde{G}_{\mu}^a (x)>$, with the effective
gluonic background field

\begin{eqnarray}
\left< \tilde{G}_{\mu}^a (x) \right> &=:&
<G^{a}_{\mu}(x)>_G +
\nonumber\\
&& \mbox{\hspace{-1cm}}
 \sum_{m=1}^{\infty} \frac{(-i g)^m}{(m+1)!}
\int \dddd x_1 \dots \int \dddd x_m  j^{\mu _1}_{a_1} (x_1) \dots
 j^{\mu _m}_{a_m} (x_m)
< G^{a}_{\mu}(x)  \dots  G^{a_m}_{\mu _m}(x_m) >_c\nonumber\\
\end{eqnarray}

The corresponding effective Lagrangian

\begin{equation}
{\cal L}_{{\rm eff}}(x) = \bar{\psi}(x) \left[ i\gamma ^{\mu}
\partial_{\mu}(x) - m_0 \right] \psi(x) - g j_a^{\mu}(x)
 <\tilde{G}^a_{\mu}(x)>,
\label{lag}
\end{equation}

describes the propagation of a quark  in the presence of
a gluonic background created by other quark sources.

\medskip

In the following we restrict ourselves to lowest non-vanishing order in
the expansion $(7)$, which gives for the interaction part of eq.
$(\ref{lag})$:

\be
{\cal L}_{{\rm eff}}^{\rm int}(x) = - i \frac{g^2}{2}  j_a^{\mu}(x)
\int \dddd y j^{\nu}_b (y) \left< G_{\mu}^a (x ) G_{\nu}^b (y)\right>_c .
\label{eff}
\ee

Note that this approximate expression still incorporates important features
of non-perturbative gluon-dynamics. Its effects are twofold: first, it
dresses single quarks, shifting the current quark mass $m_0$ to (average)
constituent masses $m$. At high energies this effect is negligeable because
all masses are small compared to the center of mass energy. Secondly,
it generates interactions between the color currents of projectile and
target quarks.

\bigskip

{\bf Quark-Quark-Scattering.}
In the next step we construct the quark-quark scattering amplitude in a
high energy approximation.
Using standard LSZ reduction we write down
the corresponding $4$-point Green function
 and arrive at the scattering matrix:

\bea
\left<p_3 p_4\right|{\cal S} - 1\left|p_1p_2\right>& \equiv&
i \left( 2 \pi \right)^4 \delta^4(p_4 + p_3 -p_2 - p_1)
{\cal M}_{qq}\nonumber\\
&&\mbox{\hspace{-4.2cm}}=
\int \dddd x_3 \mspace \int \dddd x_4
\mspace \expo{i \left(x_4 \cdot p_4 + x_3 \cdot p_3 \right)}
\bar{u}(p_3) \bar{u}(p_4)
(i \no{\vec{\partial}}_{x_3} - m )
(i \no{\vec{\partial}}_{x_4} - m ) \Psi (x_3,x_4;p_1,p_2).
\nonumber\\
\label{smatrix}
\eea

The full two particle wave function $\Psi (x_3,x_4;p_1,p_2)$ is defined as:

\bea
\Psi (x_3,x_4;p_1,p_2) & =&
\int \dddd x_1 \int \dddd x_2
 \expo{-i \left(x_2 \cdot p_2 + x_1 \cdot p_1 \right)}
\left< 0 \right| T \psi (x_4) \psi (x_3)
\bar{\psi} (x_2) \bar{\psi} (x_1)
\left| 0 \right>
\nonumber\\
&&
(i \stackrel{\leftarrow}{\no{\partial}}_{x_2} + m ) u (p_2)
(i \stackrel{\leftarrow}{\no{\partial}}_{x_1} + m ) u (p_1) .
\label{Welle}
\eea

The free quark wave functions are $\phi^0_p(x) = u(p) \expo{-i p \cdot x}$,
with Dirac spinors $u(p)$.

At high energy, we are able to factorize
$\Psi (x_3,x_4;p_1,p_2)$ into a product of  single particle wave
functions, $\phi_{p_1}(x_3)$ and  $\phi_{p_2}(x_4)$,
each  moving in an external field created by the
other particle, provided that we assume the color matrix commutator
$[T_a,T_b]$ to vanish. We have convinced ourselves that corrections due to
the non-commuting part amount to at most a few percent for the leading
interaction term.
 With the approximation $(\ref{eff})$ this external potential
can be constructed  explicitly.

\medskip

To arrive at the factorization we introduce lightcone momentum variables
 $p_\pm = p^0 \pm p^3$.
We neglect backscattering and make use of the fact that
at large momenta $p$ the only non-vanishing current matrix elements
are
$< p_{\pm} |j_{\pm}^a|p_{\pm}> $ while
$< p_{\mp} |j_{\pm}^a|p_{\mp}>$ and the transverse components vanish.
The momenta of the scattered quarks differ from $p_+$ or $p_-$
only by  a small transverse
amount $|\vek{p}_\perp| <\mbox{\hspace{-4mm}}< | p|$.
Couplings to  sea quarks are suppressed in this limit.

We can now solve  the Dirac equation
for the scattered quarks in the high energy
approximation of ref. \cite{4} and  arrive at the following eikonal
expressions for the quark wave functions
$\phi_{p_1}(x_3)$ and  $\phi_{p_2}(x_4)$:

\bea
\phi_{p_1}(x) &= &
\expo{i \int_{-\infty}^{x_+} \de x_+' {\cal V}_{-}
(x_+',\vek{x}_\perp,x_-)} \phi_{p_1}^0(x)
=: \Phi_-(x) \phi_{p_1}^0(x) ,
\label{psi1}\\
&&
\nonumber\\
\phi_{p_2}(x) &= &
\expo{i \int_{-\infty}^{x_-} \de x_-'
           {\cal V}_{+}(x_+,\vek{x}_\perp,x_-')} \phi_{p_2}^0(x)
=: \Phi_+(x) \phi_{p_2}^0(x).
\label{psi2}
\eea

Here the components of  $x$ are the
lightcone variables $x_+, \vek{x}_\perp,x_-$.
The quark interacts  with the effective gluonic background field along
its lightlike path $x_+$ (or $x_-$), and  we have introduced the
operators $\Phi_-$ and $\Phi_+$ which distort the
quark wave functions along their trajectories.
The gluonic background is described by the potentials ${\cal V}_{\pm}(x)$:

\bea
{\cal V}_-(x) &=& -i \frac{g^2}{8} \mspace T_a
\int \dddd y  \mspace
 \bar{\phi}^0_{p_2}(y) \gamma_- T_b \phi_{p_2} (y)
\left< G_-^a(x) G_+^b(y) \right>_c ,
\label{potminus}\\
&&
\nonumber\\
{\cal V}_+(x)& =& -i \frac{g^2}{8} \mspace T_a
\int \dddd y  \mspace
 \bar{\phi}^0_{p_1}(y) \gamma_+ T_b \phi_{p_1} (y)
\left< G_+^a(x) G_-^b(y) \right>_c .
\label{potplus}
\eea

Each of these potentials incorporates the full wave function of the second
quark. Therefore the coupled  equations $(\ref{psi1})$ and $(\ref{psi2})$
have to be solved self-consistently.

Introducing  c.m. coordinates $1/2(x_3+x_4)$ and relative coordinates
$z=x_3-x_4$ in eq. $(\ref{smatrix})$ one obtains the scattering amplitude
${\cal M}_{qq}$. At high energy we approximate $\no{\partial} \simeq
(\gamma_+ \partial_- + \gamma_- \partial_+)$. We can then perform the
integration
along the relative coordinates $z_+$ and $z_-$, and we are left with
an integral over the transverse coordinates $\vek{b} = \vek{z}_\perp$.
The resulting quark-quark scattering amplitude is

\bea
{\cal M}_{qq}(\vek{q}^2)& = &
\frac{i}{2} \bar{u}(p_3)\gamma_+ u(p_1)\bar{u}(p_4)\gamma_- u(p_2)
\int \dd \vek{b} \mspace \expo{-i \vek{q} \cdot \vek{b}}
\nonumber\\
&&\tr \left[ \Phi_- (\infty,\vek{b}/2,0) - 1\right]
\tr \left[ \Phi_+ (0,-\vek{b}/2,\infty) - 1\right],
\nonumber\\
\label{tqq}
\eea

where a color trace has been taken.

\bigskip

{\bf Gluonic Correlation Function.} The potentials ${\cal V}_\pm(x)$
involve the gluonic correlation function
$<G_{\pm}^a(x) G_{\mp}^b(y)>_c$.
With the help of the nonabelian Stokes theorem \cite{9} we can transform the
integrations over the paths along the gluon fields into surface integrals
over the field strengths $G_{\mu\nu}^a$:

\begin{equation}
\phi (x) = {\cal S} \exp\left\{ - \frac{g^2}{4} T_a T_b \int
d\sigma^{+\perp}
\int d \sigma^{-\perp} \left< G_{-\perp}^a (x) G_{+\perp}^b (y)
\right>_c \right\}  \phi^{(0)} (x)
\end{equation}

Here ${\cal S}$ refers to surface ordering which is the remnant of the
former pathordering, $d\sigma^{\pm\perp}$ are the
surface elements.
To identify our correlator with the gauge invariant correlation function
$< g^2 G_{\mu\nu}^a (x) \phi(x,y;x_0) G_{\rho\sigma}^b (y) >_c$
 we have to work in the Fock-Schwinger-gauge, where the Schwinger string
$\phi(x,y;x_0) = 1$. At this point we loose manifest
local SU(3)-color gauge invariance
and keep only global SU(3)-color symmetries.
Note that there is also a dependence on the reference point $x_0$. This
dependence can be shifted systematically into the higher order cumulants
\cite{8}.

 The correlation function $\left< G_{\mu\nu}^a (x) G_{\rho\sigma}^b (y)
\right>_c$ is our main input.
Using Lorentz and translational invariance we can write it
as \cite{10}

\begin{equation}
\left< G_{\mu\nu}^a (x) G_{\rho\sigma}^b (y)
\right>_c = \left< G_{\mu\nu}^a (x) G^{b \mu\nu } (x)
\right>_c \left[ g_{\mu\rho} g_{\nu\sigma} - g_{\mu\sigma} g_{\nu\rho}
\right] D\left(\frac{|x-y|}{a}\right)
\label{correlator}
\end{equation}

where $D(|x-y|/a)$ is a function of the distance of the correlated fields.
The most general expression includes also a derivative term which
fulfills the abelian Bianchi identities, whereas the nonabelian
character is described by the nonderivative part \cite{8}.
In lattice calculations the nonabelian term $(\ref{correlator})$
is found to dominate \cite{6}.
For the nonlocality $D(|x-y|)/a)$ we use a parametrization
given by Dosch et al.\cite{11a}

\be
D\left( \frac{|x-y|}{a} \right) =  \int \pppp{k}
\expo{- \frac{i}{a} k \cdot (x-y)}
\left(-\frac{27 \pi^4}{4}
\frac{k^2}{\left(k^2 - 9 \pi^2 / 64\right)^4}\right),
\ee

normalized to  $D(0) =1$.
This introduces the gluonic correlation
length $a$ which enters in the calculation
of cross sections. The diagonal elements
$\left< g^2 G_{\mu\nu}^a (x) G_a^{\mu\nu} (x) \right>_c$ are
related to the gluon condensate \cite{11}. The nondiagonal
elements vanish when taking the color trace.

Note that in leading order the potentials ${\cal V}_\pm$, eqn.
($\ref{potminus},\ref{potplus}$), are proportional to the gluon condensate
times the non-locality $D$ which scales like  $a^4$.

\bigskip

{\bf Differential Cross Sections.}
Given the quark-quark scattering matrix ${\cal M}_{qq}$, we still need to know
the transverse distributions of projectile and target quarks. We describe
those by a properly normalized form factor $F(\vek{q}^2)$. We introduce
the eikonal phase, or profile function

\be
\chi (\vek{b}^2) = \int \pp{\vek{q}} \expo{i \vek{q} \cdot \vek{b} }
F^2(\vek{q}^2) {\cal M}_{qq} (\vek{q}^2)
\ee

and construct the scattering matrix for the composite hadrons:

\begin{equation}
{\cal M} (\vek{q}^2)  = 2
\int \dd \vek{b}\mbox{\hspace{1mm}} \expo{-i \vek{q} \cdot \vek{b}}
\left[\expo{i \chi (\vek{b}^2) } - 1 \right].
\label{tmatrix}
\end{equation}

The differential cross section is

\be
\frac{d\sigma}{dt} = \frac{1}{16 \pi}
\left| {\cal M}(t)\right|^2 ,
\ee

and the total cross section satisfies the optical theorem

\be
\sigma _{tot} = \mbox{Im} {\cal M} (t=0).
\ee

For the hadronic formfactor we choose a monopole form
$F(t) = \lambda^2/(\lambda^2 -t)$
with $\lambda = \sqrt{6}/r_c$, where $r_c$ plays the role of a confinement
radius. Taking a dipole or a gaussian formfactor instead
would not change the value of $r_c$ very much but would  result in a
slightly less optimal fit to the cross section at $-t > 2\GeV^2$.

\bigskip

In table $1$ we list our best-fit parameters,  which reproduce
total and differential cross sections for
$pp$ and $p\bar{p}$ scattering.
We find good agreement with the ISR $pp$ data \cite{12} as well as with the
$p\bar{p}$ data at $\sqrt{s} = 546 $GeV \cite{13}.

\bigskip

{\bf Summary and discussion}

We have presented a description of high energy elastic hadron scattering
using a model characterized by two basic scales: the gluonic correlation
length and a confinement radius. The gluon correlation length determines
the interaction of constituent quarks in the projectile with those in the
target. The strength of this interaction is governed by the gluon condensate
for which we find values remarkably close to results obtained in the
QCD sum rule approach \cite{11}.

The gluon correlation length $a$ turns out to be primarily responsible for the
dip in the differential cross section. Its value $a\simeq 0.2$ fm is small in
comparison with typical hadron sizes and agrees with recent SU(3) lattice
results on the gluon field strength correlator\cite{6}. It is also compatible
with the small size of the scalar glueball wave function extracted from
lattice calculations \cite{13a}. We note that the dip in $d\sigma/dt$
shows up in the amplitude already before eikonalization.

A confinement scale $r_c \simeq 0.5 - 0.6$ fm completes the input to
obtain a good overall fit to the measured differential cross sections for both
$pp$ and $\bar{p}p$. Both length scales, $a$ and $r_c$, together determine
the slope of  $d\sigma/dt$ at small $|t|$. This slope is frequently
interpreted in terms of effective mean square radii of the interacting hadrons:

\begin{equation}
b = \left( \frac{d}{d t} ln \frac{d \sigma}{d t} \right)_{t=0} =
\frac{1}{3} \left( <r^2>_{\rm projectile} + <r^2>_{\rm target} \right).
\label{slope}
\end{equation}

Phenomenological analysis \cite{15} emphasizes the apparent equality
of such hadronic radii with the empirical electromagnetic radii.
The present study points to a different interpretation.

Dosch et al. \cite{16}, in a similar investigation,
 give larger values for both the correlation length $a$ and the confinement
radius $r_c$.  For the latter they
take the value of the
proton  charge radius, $r_c = 0.86$ fm, and find $a=0.35$ fm.
We can reproduce their results  by calculating the slope parameter
$b$ using our scattering amplitude, but restricted to the Born approximation,
i.e. the lowest order contribution only. We point out that the full
calculation, with the basic quark-quark interaction iterated to all
orders, induces changes of roughly $30$\% in comparison with the Born
amplitude which result in the smaller values of $a$ and $r_c$ given in
table $1$.

\bigskip

\begin{center}

\begin{tabular}{|c|c|c|c|}\hline
   &  figure 1a & figure 1b & figure 2 \\
 & $pp$  & $pp$ & $p\bar{p}$ \\
$\sqrt{s} = $  &  $52.8 $GeV & $62.5 $GeV &  $ 546 $GeV \\
 \hline
        $a$ (fm)         & 0.196  & 0.199 & 0.220\\
        $r_c$ (fm)         & 0.54 & 0.54 &0.53  \\
 $<(\alpha/\pi) G^2>$ (GeV\hspace{0.5mm}$^4$) & 0.014 & 0.014 & 0.008 \\\hline
\end{tabular}

\end{center}

\bigskip

\renewcommand{\baselinestretch}{0.3}

{\small\sl Table 1: Gluonic correlation length $a$,
confinement scale $r_c$ and
gluon condensate $<(\alpha/\pi) G^2>$ used to reproduce the data
 for pp scattering at
$\sqrt{s} = 52.8$ GeV and $62.5$ GeV and $p\bar{p}$ scattering at
$\sqrt{s} = 546$ GeV
as shown in figs. 1,2.
(The "standard" value of the gluon condensate is
$<(\alpha/\pi) G^2>=0.012\pm0.003 \GeV\mbox{\hspace{0.5mm}}^4$\cite{11}.)
}

\renewcommand{\baselinestretch}{1.}

\bigskip

{\bf Acknowledgements}

One of us (U.G.) wants to thank O. Nachtmann and P.V. Landshoff
for intensive discussions and useful comments.

\vspace{1cm}

Figure 1.
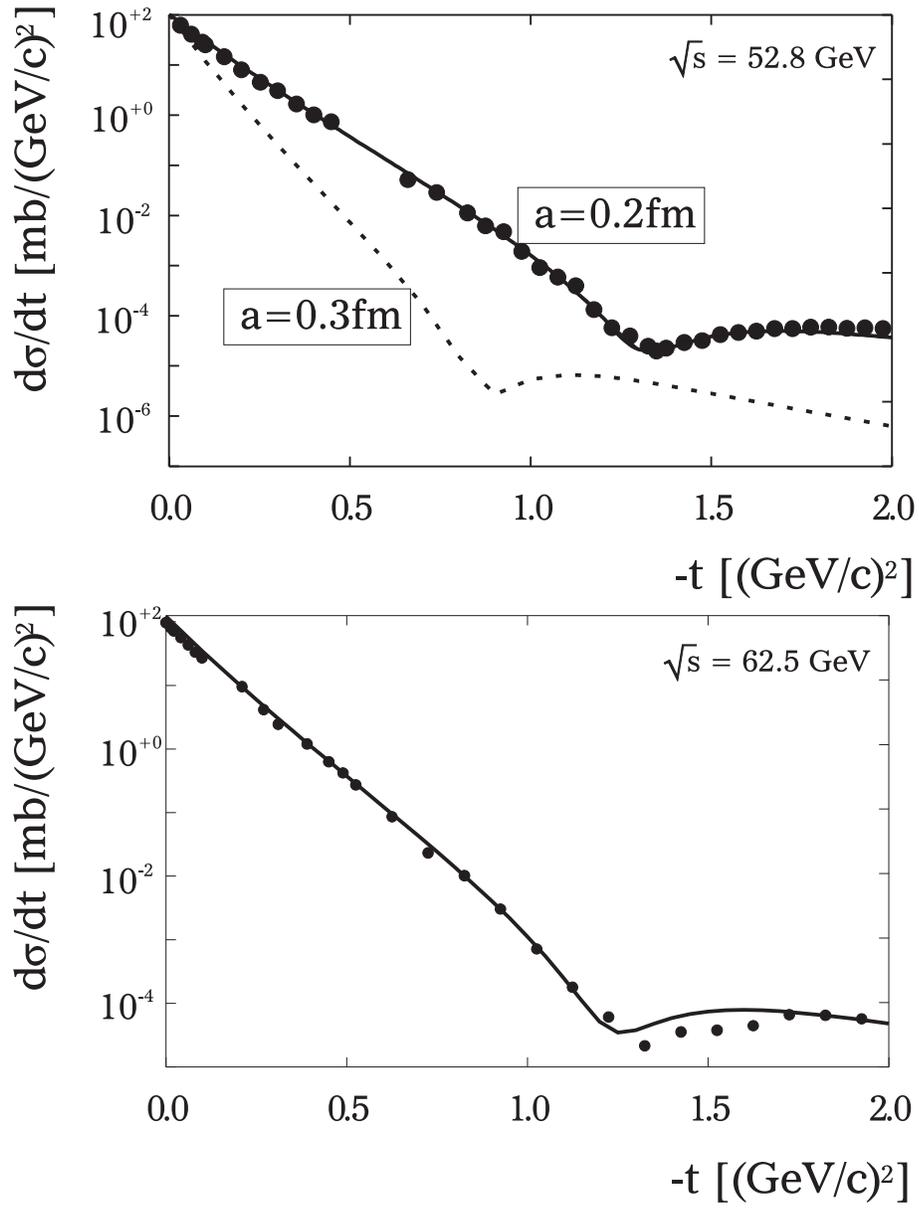
$pp$-scattering at $\sqrt{s}=52.8$ GeV (a) and $62.5$ GeV (b). The
lines present our optimal fits with the parameters given in table 1. Fig. 1a
shows in addition a fit with $a=0.3$ fm.The data are taken from \cite{12}.

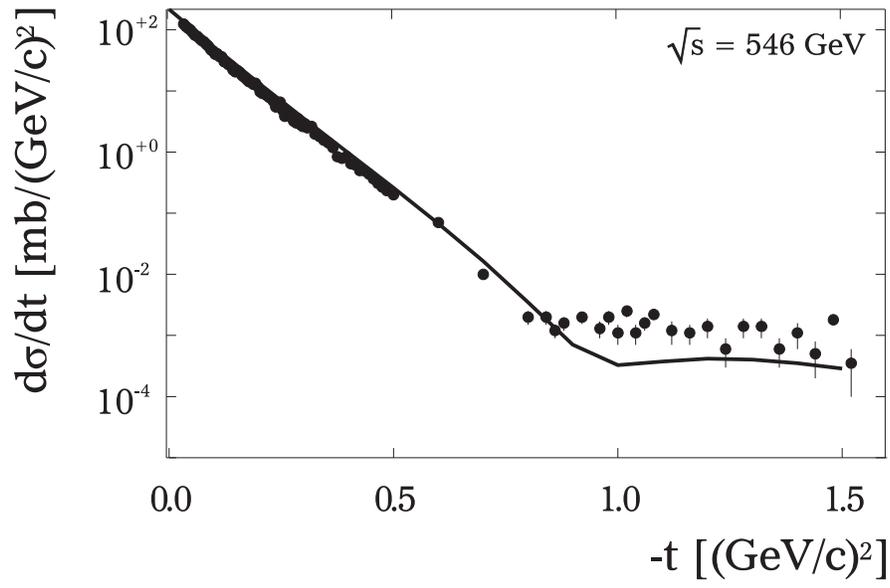
Figure 2. $p\bar{p}$-scattering at $\sqrt{s}=546$ GeV. The
line presents our optimal fit with the parameters given in table 1. The data
are taken from \cite{13}.

(Appended as uu-encoded files)

\end{document}